\documentclass{article}
\usepackage{amssymb}
\usepackage{color}
\usepackage{graphics,bm}
\usepackage{natbib}
\usepackage{amsmath}
\usepackage{xcolor}
\usepackage{latexsym}
\usepackage{adjustbox}
\usepackage{anyfontsize}
\usepackage[a4paper]{geometry}
\usepackage[textwidth=8em,textsize=small]{todonotes}
\usepackage[english]{babel}
\usepackage{times}
\usepackage{enumerate}
\usepackage[super]{nth}
\usepackage{epsfig}
\usepackage{amsfonts}
\usepackage[english]{babel}
\usepackage{graphicx}
\usepackage{amsbsy}
\usepackage{textcomp}
\usepackage{outlines}
\usepackage{etoolbox}
\usepackage{grffile}

\begin{document}
\title{Does hospital cooperation increase the quality of healthcare?}
\author{Paolo Berta\\[4pt]
	\textit{Department of Statistics and Quantitative Methods, University of Milano Bicocca,  Italy}\\[2pt]
	{paolo.berta@unimib.it}\\[4pt]
	Veronica Vinciotti\\[4pt]
	\textit{Department of Mathematics, Brunel University London, UK}\\[4pt]
	Francesco Moscone\\[4pt]
	\textit{Business School, Brunel University London, UK}}	
\date{}
\maketitle

\begin{abstract}
Motivated by reasons such as altruism, managers from different hospitals may
engage in cooperative behaviours, which shape the networked healthcare
economy. In this paper we study the determinants of hospital cooperation and
its association with the quality delivered by hospitals, using Italian
administrative data. We explore the impact on patient transfers between
hospitals (cooperation/network) of a set of demand-supply factors, as well
as distance-based centrality measures. We then use this framework to assess
how such cooperation is related to the overall quality for the hospital of
origin and of destination of the patient transfer. The over-dispersed
Poisson mixed model that we propose, inspired by the literature on social
relations models, is suitably defined to handle network data, which are
rarely used in health economics. The results show that distance plays an
important role in hospital cooperation, though there are other factors that
matter such as geographical centrality. Another empirical finding is the
existence of a positive relationship between hospital cooperation and the
overall quality of the connected hospitals. The absence of a source of
information on the quality of hospitals accessible to all providers, such as
in the form of star ratings, may prevent some hospitals to engage and
cooperate with other hospitals of potentially higher quality. This may
result in a lower degree of cooperation among hospitals and a reduction in
quality overall.
\end{abstract}

Keywords: hospital cooperation, patient flows, social relation model,
healthcare quality

\section{Introduction}

\label{sec:Introduction} In recent years, several central and local
governments in Western countries such as the UK and Italy, have implemented
pro-competition reforms in their healthcare sectors with the view that, as
predicted by the economic theory, more competition among hospitals, when
prices are regulated, would lead to improvements in the quality of healthcare services, ultimately having a positive impact on the health outcomes of
the population. While there has been a wide and alive debate among health
economists on the effects of competition on hospital quality \citep{berta2016association,chone2017competition,colla2016hospital,mascia2017,mukamel2002hospital,NBERw12301,gaynor2017making,propper2004does}, little is known on cooperative behaviours among healthcare providers, and
their effects on health outcomes. The question whether cooperation exists is
important. If it does, we first need to understand the mechanisms underlying
cooperation and then whether it has a positive, a null or even a negative
effect on hospital quality.

Why should economic agents cooperate? Economists have addressed this
question with game theory, starting with the example of the prisoner dilemma
to more complex dynamic games that find their applications in different
areas of applied economics such as trade. Similarly, in the healthcare
sector doctors, economic actors such as nurses, or managers may decide under
certain rules, strategies, and payoffs, to cooperate rather than to act
independently. Managers from different hospitals, who are motivated by
reasons such as convenience or altruism, may decide to cooperate, with the
aim of improving efficiency and health outcomes of their respective
organizations \citep{Gittell2004,mascia2012dynamic}. Such cooperation can
take various forms, ranging from merging facilities to clinical network
information sharing, joint treatment or joint diagnostic centres, new shared
assets and joint construction of new facilities. Informal cooperation
between healthcare providers may also take place. This occurs for example
when we observe the existence of a network between professionals, healthcare providers or the management boards of different hospitals \citep{westra2017understanding}. For example, physicians from any two
hospitals may collaborate when treating a patient, thus creating
correlations in health indicators across hospitals \citep{westra2016}.
However, these networks tend to predominate within rather than between
hospitals \citep{barnett2011mapping,landon2012variation,pollack2012physician}.

In this paper, we study the informal network that is generated among healthcare providers when a patient is transferred across hospitals. Under the
assumption that hospital managers are altruistic agents, hospitals decide to
transfer patients to other hospitals when the benefits of transfer outweigh
the risks. While the decision to transfer a patient is usually driven by the
availability of specialized care in the hospital of origin and destination,
the choice of the destination hospital may be driven, among other things, by
its geographical proximity, demand-supply factors of the hospital of origin
and destination, as well as the relative quality of the hospital of origin
with respect the hospital of destination. However, hospital managers in some
countries may not know the distribution of quality across the other
hospitals in the healthcare sector, thus their choice will be driven by a
measure of perceived relative quality. This source of asymmetric information
may produce different effects: if the relative perceived quality is
reflecting the relative ``true quality",
we should expect that cooperation will improve overall health outcomes of
both hospital of origin and destination. However, if the relative perceived
quality is negatively associated with the relative ``true
quality", cooperation may even harm patients in both
hospitals. Policy makers may have a strong interest in understanding the
drivers underlying these cooperation networks in order to be able to design
effective policy interventions, as well as identifying healthcare provider
links that actually improve the health outcome of the transferred patient.

In this paper we study the determinants of hospital cooperation and its
association with the quality delivered by the networking hospitals. We first
explore the impact on patient transfers between hospitals (cooperation
network) of a set of demand-supply factors, as well as centrality measures
from the network of geographical distances, including factors related to
hospital quality. The decision of including quality in the manager's
hospital transfer choice is coherent with our context whereby policy makers
know the distribution of adjusted quality across the territory, and this
information is partially revealed to the hospital managers, as explained
further in Section \ref{sec:data}. On the other hand, in order to study the
effects of cooperation (patient transfer) on overall quality for the
hospital of origin and of destination, the first stage will be estimated after
excluding hospital quality and all the variables possibly correlated with it,
since real patient transfer flows can be influenced by hospital quality if
the decision of the referring hospital is based, among other things, on the
relative quality of the destination hospital. Such endogeneity may bias
results when regressing the transfers on health outcomes.

Following the literature on social relations models \citep{warner1979new,hoff2005bilinear}, we adopt an over-dispersed Poisson
mixed model that is suitable to handle network data. These statistical
models are rarely used in health economics and health research in general.
We use data on hospital discharges for over 900,000 patients admitted to 145
hospitals in the Lombardy region (Italy) in 2014. Among these patients,
around 15,500 (1.7\%) were transferred to other hospitals after admission.
Our results show that geographical distance plays an important role in
hospital cooperation, although there are also other factors that matter,
such as the geographical centrality of a hospital. Another empirical finding
is the existence of a positive relationship between hospital cooperation and
the overall clinical quality for the hospital of origin and of destination.

The remainder of the paper is organized as follows. Section \ref{sec:bckgr} reviews the
literature on the determinants of patient flows and the impact of
cooperation on hospital quality. Section \ref{sec:data} describes the data, introducing
the Italian NHS and the Lombardy healthcare system, which is the focus of
our empirical investigation. Section \ref{sec:net} undertakes an exploratory data
analysis of networks of transfers. Section \ref{sec:mod1} estimates patient flows via an
over-dispersed Poisson mixed models. Section \ref{sec:ES} estimates the impact of
cooperation on the hospital quality. Section \ref{sec:Discussion} makes some concluding remarks and plans for future work.

\section{Literature background}

\label{sec:bckgr}

The basis for modelling patient flows across geographical locations is the
gravity model \citep{silva2006log}, which involves a ``mass'' term for both
the origin and destination units, and incorporates the impact of
geographical distance. There exists a strand of literature in health
economics that uses the gravity model to investigate the determinants of
patient flows at regional, Local Health Authority (LHA) or hospital/ward
level \citep{levaggi2004patients,shinjo2012geographic,congdon2001,balia2018interregional,cantarero2006health,fabbri2010geography,mascia2012dynamic}. For example, \cite{balia2018interregional} adopt a gravity model to
investigate the determinants of patient mobility among Italian regions for
the period 2001-2010 using data on hospital discharges. The authors find
that income, hospital capacity and the regional technological level are the key
drivers of patient regional flows. \cite{congdon2001}, using data on
emergency units in 127 electoral wards in North East London and Essex, finds
patient age and the travel distance to be the main drivers to patient flows.
In general, regardless of the level of aggregation of the data used, these
studies find that the most important variables explaining patient transfers
are: geographical distance between healthcare providers, patient
characteristics, the capacity of hospitals, the availability of medical
technologies, and the quality of health services.

There is an emerging interest in studying the determinants of patient
hospital transfers by adopting a network analysis approach. Some authors have
summarised the network in the form of centrality measures, some others have
modelled directly the presence of an edge between two hospitals, i.e. a
cooperation between two hospitals. In particular, \cite{lomi2012relational}
and \cite{caimo2017bayesian} use exponential random graph models to link the
tendency of hospitals to cooperate with hospital characteristics and a set
of network summary statistics, such as the density of the network, the
presence of mutual edges, or reciprocity, and of triads. Using
administrative data for 91 hospitals located in the Lazio region (Italy),
the authors find that hospitals' proximity and sharing an administrative
membership facilitate cooperation. Furthermore, they find the presence of
local networks, with the tendency to reciprocity among hospitals.

Differently from the works cited above, \cite{mascia2015effect} studies the
effect of patient transfers, and particularly of the topological properties
of the network of flows, on hospital quality. The authors adopt a multilevel
model approach to describe the impact of measures of centrality and
ego-network density of the network of transfers on readmissions within
45-days after the discharge. Using administrative data for 31 hospitals in
the region of Abruzzo (Italy), they find that greater network centrality, in
the form of hospitals with many flows with other hospitals who are also
central, is negatively associated with readmissions, whereas greater
ego-network density, represented by a high sharing of patients among
hospitals that are connected to a central hospital, increases the likelihood
of readmissions, thus reducing the quality provided.

The works reviewed above have contributed to identifying the determinants of
cooperation between hospitals, and will be used in Section \ref{sec:mod1} as
a guidance to specify our empirical model. In particular, we will model
cooperation between hospitals by including demand-supply variables, hospital
quality, geographical distance as well as distance-based centrality measures
that account for spatial correlation in the data. Further, similarly to \cite{mascia2015effect}, we will consider mortality rate as well as readmissions as health outcomes, but we perform the analysis on a different Italian region (Lombardy) and across a
larger number of hospitals (145). In contrast to \cite{mascia2015effect}, however, we
also consider the fact that quality may be a contributing factor in the
decision of a hospital to transfer a patient to a different hospital, and we
therefore opt for a two-stage approach.

\section{Data}

\label{sec:data}

The Italian National Healthcare System (NHS) follows the
Beveridge model \citep{beveridge1942social}, providing universal healthcare
coverage throughout the country as a single payer. It entitles all citizens,
regardless of their social status, to equal access to essential healthcare
services. In 1992, a system reform transferred administrative and
organizational responsibilities and tasks from the central government to the
administrations of the 21 regions in Italy. These regions now have
significant autonomy on the revenue side and in organizing services designed
to meet the needs of their respective populations.

The Lombardy healthcare system was reformed in 1997 becoming a quasi-market
system made up of both public and private providers which are reimbursed by
a prospective payment system based on Diagnosis Related Groups (DRGs) \citep{brenna2011quasi,berta2010effetcs,berta2013comparing}. The
reimbursement provided to the hospitals for each discharge is defined
according to specific DRG tariffs, revised every year by the regional
government on the basis of increasing costs due to the introduction of new
medical technologies and also taking into account the introduction of new
policies. These public regional reimbursements represent the majority of
revenues for acute discharges in all hospitals located in Lombardy. The
Lombardy healthcare system yearly provides the results on hospital quality
on a web portal in which hospitals can access and see their performance
rankings with respect to other hospitals within the region. The regional
health authority provides a hospital classification into three groups
depending on whether the quality is significantly above, not different, or
significantly below the regional average performance. Along this
information, managers may hold informal information on the quality of
hospitals; this is the reason of why in Section 5 we will control for
hospital quality when studying the impact on patient transfers between
hospitals (cooperation/network).

In this paper, we analyse data gathered from the administrative regional
healthcare information system, which includes information on patients
discharged from 145 hospitals accredited with the regional healthcare system
in the Lombardy region (Italy) in the year 2014. The dataset contains
1,541,996 hospitalizations, of which 84\% were ordinary and 16\% were in day
hospital or day surgery. Furthermore, hospitalizations of patients living
outside the Lombardy region accounted for 10\% of all admissions. The
hospital discharge data contains demographic information such as age and
gender, information on hospitalization (length of stay, special-care unit
use, transfers within the same hospital or through other facilities, and
within-hospital mortality), and a total of 6 diagnosis codes and surgical
procedures defined according to the International Classification of
Diseases, Ninth Revision, Clinical Modification (ICD-9-CM). Only ordinary
hospitalizations for patients aged more than 2 years were retained in the
sample. We define a transfer between hospitals by a patient discharged from
a hospital and then admitted in another hospital on the same day or the next
one \citep{iwashyna2009b}. In order to exclude any patient involvement in
this process, we exclude voluntary discharges. Finally, we define mortality
by the death of the patient in hospital, or within 30 days after the
discharge, and we define readmission by a patient readmission within 45-days
after the discharge and for the same major diagnostic class.
\begin{table}
\caption{\label{tab:tab0}Descriptive statistics of the patient and hospital data, for the variables that will be used in subsequent models. Information is split by hospital ownership.}
    \begin{tabular}{l|rr|rr|rr}
    \hline\hline
          & \multicolumn{2}{|c|}{Private Hospitals} & \multicolumn{2}{|c|}{Public Hospitals} & \multicolumn{2}{|c}{Overall}  \\
          & \multicolumn{1}{|c}{mean} &  \multicolumn{1}{c|}{sd} & \multicolumn{1}{|c}{mean} &  \multicolumn{1}{c|}{sd} & \multicolumn{1}{|c}{mean} &  \multicolumn{1}{c}{sd} \\
          \hline
   \textit{Outcomes}  &   &   &   &   &      \\
    Mortality rate & 0.04  & 0.04  & 0.07  & 0.04  & 0.07 & 0.04\\
    Readmission rate & 0.08  & 0.03  & 0.11  & 0.05 & 0.10 &  0.04\\
    &   &   &   &   &   &   \\
 \textit{Patient Characteristics }   &   &   &   &   &   &   \\
Female (F)	 & 0.51	 & 0.08	 & 0.56	 & 0.10	 & 0.54	 & 0.10	 \\
Age	(A) & 64.38	 & 7.00	 & 61.03	 & 7.41	 & 62.37	 & 7.41	 \\
DRG Weight (DW)	 & 1.24	 & 0.27	 &  1.10	 & 0.21	 & 1.16	 & 0.24	 \\
    &   &   &   &   &   &   \\
\textit{Hospitals Characteristics}  &   &   &   &   &      \\
Beds Saturation (BS)	 & 61.12	 & 20.55	 & 79.24	 & 10.30	 & 71.99	 & 17.60	 \\
Beds Turnover (BT) & 40.10	 & 11.43	 & 40.98	 & 8.25	 & 40.63	 & 9.62	 \\
Distance in minutes (D)	 & 59.60	 & 15.89  & 	66.83	 & 21.43	 & 63.94	 & 19.67	 \\
\hline
\# Hospitals &  \multicolumn{2}{|c|}{58} & \multicolumn{2}{|c|}{87} & \multicolumn{2}{|c}{145}	 \\
\# Transfers &  \multicolumn{2}{|c|}{3,024} 	&  \multicolumn{2}{|c|}{12,492} &  \multicolumn{2}{|c}{15,516} 	\\
\# Hospital Discharges (HD) &  \multicolumn{2}{|c|}{256,909} 	&  \multicolumn{2}{|c|}{643,242}  &  \multicolumn{2}{|c}{900,151}  \\
    \hline
    \end{tabular}%
\end{table}%
Table \ref{tab:tab0} provides a set of descriptive statistics on health
outcomes and patient and hospital characteristics, split by hospital
ownership (private and public). Around 45\% of the hospitals are private,
although they only cover 28\% of the hospitalizations. It is interesting to
observe that, while patient demographic characteristics (age and gender) are
similar for private and public hospitals, their case-mix is quite different,
with private hospitals having a higher DRG weight. In terms of health
outcomes, we observe that gross rates are higher in public hospitals
compared to the private, with a small difference for readmissions and a
bigger gap for mortality. This is not directly related with the quality
provided by the two types of providers, but it is more related with the
different case-mix of patients admitted in public and private hospitals. In
terms of hospital characteristics, the table reports the beds saturation
index, or occupancy rate, which is measured by the average number of days
when a hospital bed is occupied as a percentage of the available 365 days, and the
beds turnover index, which is a measure of the extent of beds' utilization
and is measured by the number of changes in bed occupancy during the year.
The statistics show that, while public hospitals are on average bigger than
private ones and with a higher saturation index, the turnover index is
similar between private and public hospitals. Finally, looking at the
distance between hospitals, which is measured by the travel time between any
two hospitals, the table shows how public hospitals tend to be on average
slightly more distant (approximately 7 minutes more) than private ones.
These variables will be used in the models presented in the next sections.

\section{Exploratory analysis of the network of transfers}

\label{sec:net}

As discussed in the introduction, we measure hospital
cooperation using the network of patient transfers. In this network each
hospital in Lombardy is a node and the edges are the connections between two
hospitals with a weight defined by the number of patient transfers between
the two hospitals. Table \ref{tab:tab0} reports the total number of
transfers, split by hospital ownership. In Table \ref{tab:tabnet} we further
describe the network of transfers using various measures of network
centrality from the network modelling literature \citep{Kolaczyk14}, but
also considered in the literature of patient flows \citep{fernandez2017influence}.
\begin{table}
\caption{\label{tab:tabnet}Descriptive statistics of the network of patient flows.}
  \centering
    \begin{tabular}{lrcc}
        \hline    \hline
          & Overall & Private Hospitals & Public Hospitals \\
              \hline
In-Centrality  &    5.0633 & 2.4328  & 6.3969 \\
In-Closeness    &   0.0976  & 0.0299  & 0.2061 \\
In-Strength     &   0.4454  & 0.3771  & 0.4053 \\
Betweenness    &    0.1390  & 0.1499  & 0.1385 \\
Out-Centrality &    2.9730  & 1.6258  & 3.3388 \\
Out-Closeness  &    0.0249  & 0.0170  & 0.2033 \\
Out-Strength   &    0.3273  & 0.2017  & 0.4053 \\
           \hline
\# Hospitals & \multicolumn{1}{c}{58} & \multicolumn{1}{c}{87} & \multicolumn{1}{c}{145} \\
        \hline    \hline
    \end{tabular}%
\end{table}%
In particular we consider the in-centrality of the network, which is based
on the in-degrees of all nodes in the network, i.e. the number of hospitals
from which a given hospital receives transferred patients. Similarly, the
out-centrality measures the number of hospitals to which patients are
transferred from a given hospital, across all nodes. The in- and
out-strength indices further consider the weights associated to each edge,
i.e. the number of patients received/transferred by a hospital. In addition,
we calculate the closeness of the network, indicating the proximity of each
node with the other nodes in terms of the number of steps needed to go from
a hospital to the transferring/receiving hospital. Similarly, betweenness
quantifies the number of times a hospital is a bridge for the other
hospitals in the network. Each measure is calculated using the \texttt{centr\_degree} function in the \texttt{igraph} R package \citep{csardi06},
from which normalized scores are automatically calculated which are
comparable across graphs of different sizes. The results are presented in
Table \ref{tab:tabnet} for the overall network, as well as for the two
sub-networks of private-only and public-only hospitals. Overall, the results
appear to show how the network is rather centralized and how there is a
higher tendency for nodes to have a large in-degree (i.e. receiving from
many other hospitals) than out-degree. In addition, it appears that the
network of public hospitals has more central nodes/hubs than the network of
private hospitals.

\begin{figure}[tbp]
\centering
\includegraphics[width=1\textwidth]{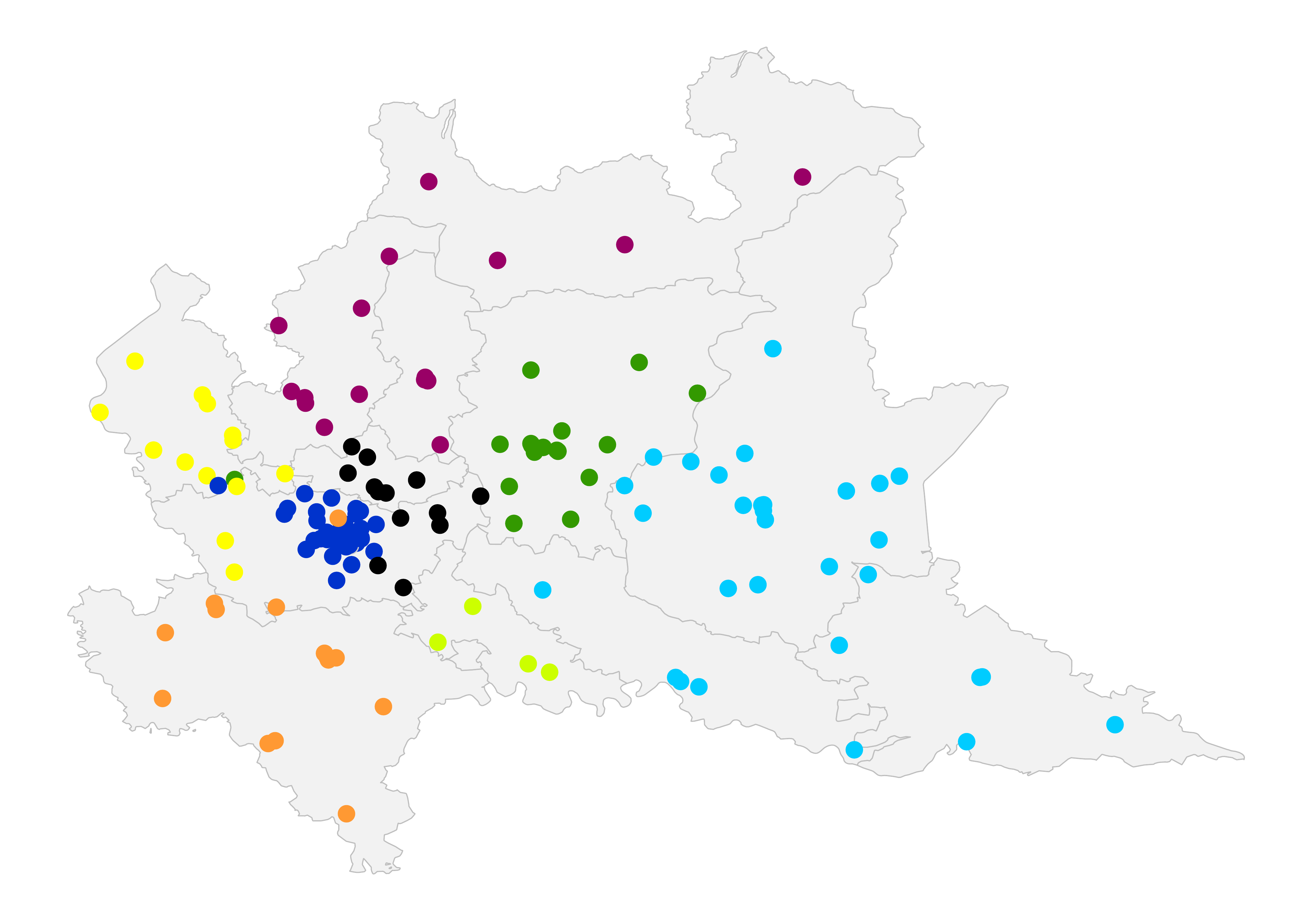}
\caption{Maps of the hospitals in Lombardy and their belonging to the
communities detected based on modularity. Each of the 9 detected communities
is assigned a different colour. The map also shows the borders between
different municipalities. }
\label{fig0}
\end{figure}

Aside from centrality measures and with a view to finding determinants of
patient flows, we next consider approaches to identify possible structures
in the network, for example in the form of partitions of the network where
nodes belonging to the same partition are strongly connected among them and
sparsely connected with the nodes belonging to different partitions. To this
aim, the network of transfers has been analyzed using a community detection
method. In particular we consider the method by \cite{Blondel2008}, which is
highlighted by \cite{yang2016comparative} in their comparative study and
where partitions are searched based on improvements of the modularity score.
Figure \ref{fig0} shows the optimal partition for the network of transfers,
obtained using the \texttt{multilevel.community} function in the \texttt{%
igraph} R package \citep{csardi06}. The figure shows a strong relationship
between the modularity-detected communities and their geographical location.
Nine communities are detected and most of them correspond to specific
municipalities in Lombardy (e.g. Pavia, Lodi and Bergamo correspond to the
orange, red and green communities, respectively ). The metropolitan area of
Milan, where the most part of the hospitals are located, is characterized by
three different communities: the eastern part of Milan shares the community
with the municipality of Monza-Brianza (black dots), the western part shares
the hospitals with the community of Varese (yellow dots), whereas the
central part identifies one single community (blue dots). Finally two more
communities are identified: one that is characterized by the presence of the
mountains in the north and include the municipalities of Sondrio, Como and
Lecco (purple dots) and the second one which is characterized by the eastern
part of the region and is shared by the municipality of Cremona, Mantua and
Brescia (azure dots). The community detection approach has shown how both
the distance between hospitals and the belonging to the same Local Health
Authorities (LHS), are substantial factors in defining the network of
hospital transfers. This finding is consistent with the literature %
\citep{lomi2012relational,mascia2012dynamic,caimo2017bayesian}. On the basis
of this explorative analysis, we will include both distance and
co-memberships to LHAs as determinants of patient transfers in the next
analysis. For that purpose, it is also important to emphasize that these
determinants are exogenous to the quality of hospitals.

\section{Modelling patient flows via an over-dispersed Poisson mixed model}

\label{sec:mod1}

In order to predict the network of patient transfers from a
number of exogenous determinants, we resort to the literature on social
relations models \citep{warner1979new}, where the statistical dependencies
inherent in dyadic data are carefully considered and accounted for. This
results in a general class of mixed effect models, where random effects are
included in order to control for node effects, which are typical of network
data where multiple observations relate to the same node/edge. These models,
not considered in the healthcare literature before, but known in the network
modelling literature \citep{hoff2005bilinear}, have been implemented in the
\texttt{amen} R package using a Bayesian inferential approach %
\citep{amen2017}. Since the dependent variable in our study is discrete, we
will extend the existing implementations to the case of an over-dispersed
Poisson mixed model.

Going to the details of the model, let $T_{ij}$ be the number of transfers
between hospital $i$ and $j$. Since the data is in the form of a network,
statistical dependencies are to be expected, for example the fact that data
associated to the same hospital of origin $i$ (i.e. a row of the matrix of
transfers) or to the same destination hospital $j$ (i.e. a column of the
matrix of transfers) may be more similar to each other than to the rest of
the observations. We model these dependencies using the mixed model
\begin{align}
E(T_{ij}|\cdot) = & \exp( \alpha + \beta_1 HD_{i} + \beta_2 HD_{j} + \zeta_1
DW_{i} + \zeta_2 DW_{j} + \gamma_1 A_{i} + \gamma_2 A_{j} + \delta_1 F_{i} + \delta_2 F_{j} +  \notag \\
& + \eta_1 DC_{i} + \eta_2 DC_{j} + \theta_1 BW_{i} + \theta_2 BW_{j} + \vartheta_1 Teach_{i} + \vartheta_2 Teach_{j} +  \notag \\
& + \lambda_1 Mono_{i} + \lambda_2 Mono_{j} + \varrho_1 Techno_{i} + \varrho_2 Techno_{j} +  \notag \\
& + \kappa_1 BS_{i} + \kappa_2 BS_{j}+ \tau_1 BT_{i} + \tau_2 BT_{j} + \omega_1 AM_{j} + \omega_2 AM_{j} + \phi_1 AR_{j} + \phi_2 AR_{j} +
\notag \\
& + \xi D_{ij} + \psi CM_{ij}+ \varepsilon_{ij}),  \label{eq7}
\end{align}
where the errors have the following dependency structure
\begin{align}
\varepsilon_{ij} = & a_i + b_j + \nu_{ij}  \notag \\
(a_{i}, b_{i} )^{^{\prime }} \sim & MVN(0,\Sigma_{ab}) \text{,}\quad
\Sigma_{ab} = \left(
\begin{matrix}
\sigma_{a}^2 & \sigma_{ab} \\
\sigma_{ab} & \sigma_{b}^2%
\end{matrix}%
\right),  \notag \\
(\nu_{ij}, \nu_{ji} )^{^{\prime }} \sim & MVN(0,\Sigma_{\nu}) \text{,}\quad
\Sigma_{\nu} = \sigma_{\nu}^2 \left(
\begin{matrix}
1 & \rho \\
\rho & 1%
\end{matrix}%
\right),  \notag
\end{align}
with $a_i$ and $b_j$, $i,j = 1,\ldots,N$, the random effects for the sender
and receiver hospitals, respectively, and $\nu_{ij}$ the errors. This model
induces a covariance among the $\varepsilon_{ij}$ given by:
\begin{equation*}
E(\varepsilon_{ij}^2) = \sigma_{a}^2 + \sigma_{b}^2 + \sigma_{\nu}^2 \text{,}%
\quad E(\varepsilon_{ij}\varepsilon_{ik}) = \sigma_{a}^2,
\end{equation*}
\begin{equation*}
E(\varepsilon_{ij} \varepsilon_{ji}) = \rho \sigma_{\nu}^2 + 2 \sigma_{ab}
\text{,}\quad E(\varepsilon_{ij} \varepsilon_{kj}) = \sigma_{b}^2,
\end{equation*}
\begin{equation*}
E(\varepsilon_{ij} \varepsilon_{kl}) = 0 \text{,}\quad E(\varepsilon_{ij}
\varepsilon_{ki}) = \sigma_{ab}.
\end{equation*}
That is, $\sigma_{a}^2$ represents the correlation of observations having a
common hospital sender, whereas $\sigma_{b}^2$ defines the dependence of
observations having a common hospital receiver. Since the network is
asymmetric, $\rho$ measures the ``reciprocity'' between sender and receiver
hospitals, that is the dyadic correlation between the number of transfers
from $i$ to $j$ and those from $j$ to $i$. In addition, $\sigma_{\nu}^2$
accounts for over-dispersion: when $\sigma_{\nu}^2$ is zero, the model is a
simple Poisson mixed model, whereas when $\sigma_{\nu}^2$ increases, the
conditional variance of $T_{ij}$ becomes larger than the mean.

Several covariates are included in the model in Equation \eqref{eq7}, both
at the node and at the dyadic level. In line with the literature \citep{mascia2012dynamic,mascia2015effect} we include in the model the
dyadic covariates of geographical distance between two hospitals ($D$) and
their co-membership (CM), an indicator variable identifying if the two
hospitals belong to the same LHA. We also control for the degree centrality (%
$DC$) of both origin ($i$) and destination ($j$) hospital. This is measured
based on a geographical network where two hospitals are linked if they are
less than 30 minutes of effective time travel apart. This variable allows to
adjust the predictions for the hospitals' concentration in a pre-defined
space, the hypothesis being that a higher value of this index for the origin
hospital indicates a wider choice set for the hospital that needs to decide
where to transfer a patient, and similarly for the destination hospital. We
also include in the model the betweenness index ($BW$), calculated for each
node of the geographical network and rescaled to a minimum of zero and a
maximum of one. In terms of hospital or node-based characteristics, we
control for the number of discharges of both the sender and receiver
hospitals ($HD$) as well as for the severity of patients treated in a
hospital. For the latter, we include both the variable DRG Weight ($DW$),
taken also as a measure of resources that the hospital employs to treat
patients \citep{berta2013comparing}, and the patient age ($A$), measured as
an average at the hospital level. Similarly, we control for the patient
hospital mix in terms of gender composition by including in the model the
percentage of female patients in a hospital ($F$). Furthermore, we include
specific hospital characteristics for both sender and receiver hospitals
which identify if the hospital is a teaching hospital or not ($Teach$), if it
is a monospecialized or a general hospital ($Mono$) and if the hospital is
highly equipped or not in terms of technology ($Techno$). Finally, we include
hospital beds saturation and hospital beds turnover indices, in order to
measure the beds capacity of a hospital and the efficiency in using beds %
\citep{lomi2014quality,mascia2017}, and two variables measuring the hospital quality: adjusted mortality ($AM$) and adjusted 45-days readmissions ($AR$).

We estimate the model in Equation \ref{eq7} via a Bayesian MCMC algorithm,
adapting the implementation in the \texttt{ame} function of the \texttt{amen}
R-package \citep{amen2017} to the case of a Poisson distributed dependent
variable. We use a burn-in window of 1000 iterations, followed by 10000
iterations, where we save the estimated parameters every 25th iteration.
Table \ref{tab:1stage} shows the posterior mean estimates of the parameters.

The variable geographical distance shows that the shorter the distance
between two hospitals the higher the number of transfers. This was to be
expected from our earlier exploratory analysis (Figure \ref{fig0}), despite
the high density of hospitals in Lombardy. Similarly, as expected, the
co-membership of the hospitals to the same LHA increases the patient flows
between hospitals. These associations are reported also by other studies %
\citep{caimo2017bayesian,landon2012variation,mascia2012dynamic} and
explained by the fact that low distance in the transfer of a patient is
typically prefered by hospitals in order to reduce the costs for travelling
and the risks for the patient associated to the transfers. The positive and
significant relationship between the degree centrality and the transfers
indicates that when the set of opportunities for transferring/receiving
patients increases, the hospitals tend to transfer/receive more patients. In
order to check the robustness of this result, we have repeated the analysis
using degree centrality indices based on several thresholds of the distances
(between 20 and 40 minutes). We have observed a correlation of over 0.90
among the predicted values across these different thresholds, suggesting a
robustness of this finding. Further, adjusting for hospital discharges
proves to be important when explaining variation in patient flows.

With regards to patient characteristics, the analysis suggests that the demographic characteristics of the
average patient in a hospital, such as age and gender, do not have a significant impact on the number of patient transfers, whereas the severity
of their conditions and the ability of the hospitals in managing beds have a significant impact on the number of patient transfers. In terms of other hospital characteristics, we find that being a destination
teaching hospital increases the transfers, and, similarly, being a
technological hospital is positively related with the transfers for both
origin and destination hospitals. This might suggest that hospitals with a
low level of technology move complicated patients to the technological
hospital in order to provide for example a cardiosurgical intervention, but
when the patients overcome the critical post-surgical phase, they are likely to
be moved back to the previous hospital. In this way, the patient receives necessary
assistance, releasing the bed in the technological hospital for a new
hospitalization. Finally, the variables attached to hospital quality do
not seem to play a role in explaining patient flows, most likely due to the fact that hospital managers in the Lombardy Region do not fully know the distribution of quality across the healthcare system.

The inference also provides a quantification of the level of dependencies in
the data: in particular, the relatively large value of the parameter $\rho $
suggests a high correlation of the two observations associated to an edge in
the network and thus a high level of reciprocity between connected
hospitals. The same can be said for $\sigma _{a}^{2}$ and $\sigma _{b}^{2}$,
whose large values suggest the presence of row and column effects. These
findings support the need for the use of more advanced mixed effect models
in our study.

\begin{table}
\caption{Posterior means of estimates and standard deviations for the model
of patient transfers in Equation \eqref{eq7}. Signficance *** $0.01$, ** $%
0.05$, * $0.1$ }
\label{tab:1stage}\centering
\begin{tabular}{lr@{\hskip .001in}lr}
\hline\hline
& Estimate &  & Std. Error \\
(Intercept) & -12.931 & *** & 3.147 \\
Distance & -0.070 & *** & 0.002 \\
Co-membership & 1.787 & *** & 0.081 \\
&  &  &  \\
\textit{Origin} &  &  &  \\
Hospital Discharges & 0.075 & *** & 0.016 \\
DRG Weight & -0.012 &  & 0.442 \\
Age & 0.026 &  & 0.019 \\
Female & -0.804 &  & 1.053 \\
Degree Centrality & 0.059 & *** & 0.004 \\
Betweenness & -0.678 &  & 0.567 \\
Teaching Hospital & 0.075 &  & 0.222 \\
Monospecialized Hospital & -0.373 &  & 0.290 \\
Technological Hospital & 0.887 & *** & 0.203 \\
Public vs Private & 0.681 & ** & 0.209 \\
Risk-Adj. Mortality & -0.035 &  & 0.025 \\
Risk-Adj. 45-days Readmissions & 0.033 &  & 0.028 \\
Beds Saturation & 0.023 & *** & 0.006 \\
Beds Turnover & 0.009 &  & 0.011 \\
&  &  &  \\
\textit{Destination} &  &  &  \\
Hospital Discharges & 0.088 & *** & 0.017 \\
DRG Weight & 1.199 & ** & 0.474 \\
Age & -0.035 & * & 0.019 \\
Female & -1.831 & * & 1.106 \\
Degree Centrality & 0.046 & *** & 0.004 \\
Betweenness & 0.449 &  & 0.580 \\
Teaching Hospital & 0.536 & ** & 0.230 \\
Monospecialized Hospital & 0.328 &  & 0.312 \\
Technological Hospital & 0.918 & *** & 0.205 \\
Public vs Private & 0.288 &  & 0.213 \\
Risk-Adj. Mortality & 0.034 &  & 0.027 \\
Risk-Adj. 45-days Readmissions & 0.026 &  & 0.028 \\
Beds Saturation & 0.026 & *** & 0.007 \\
Beds Turnover & -0.003 &  & 0.011 \\ \hline
$\sigma_{a}^2$ & 0.544 &  & 0.102 \\
$\sigma_{ab}$ & 0.362 &  & 0.086 \\
$\sigma_{b}^2$ & 0.616 &  & 0.114 \\
$\sigma_{\nu}^2$ & 1.961 &  & 0.097 \\
$\rho$ & 0.886 &  & 0.016 \\ \hline\hline
\end{tabular}%
\end{table}

\begin{figure}[tbp]
\caption{Posterior predictive goodness of fit for the model in equation
\eqref{eq7}. }
\label{Gof1_1}\centering
\includegraphics[scale = 0.3]{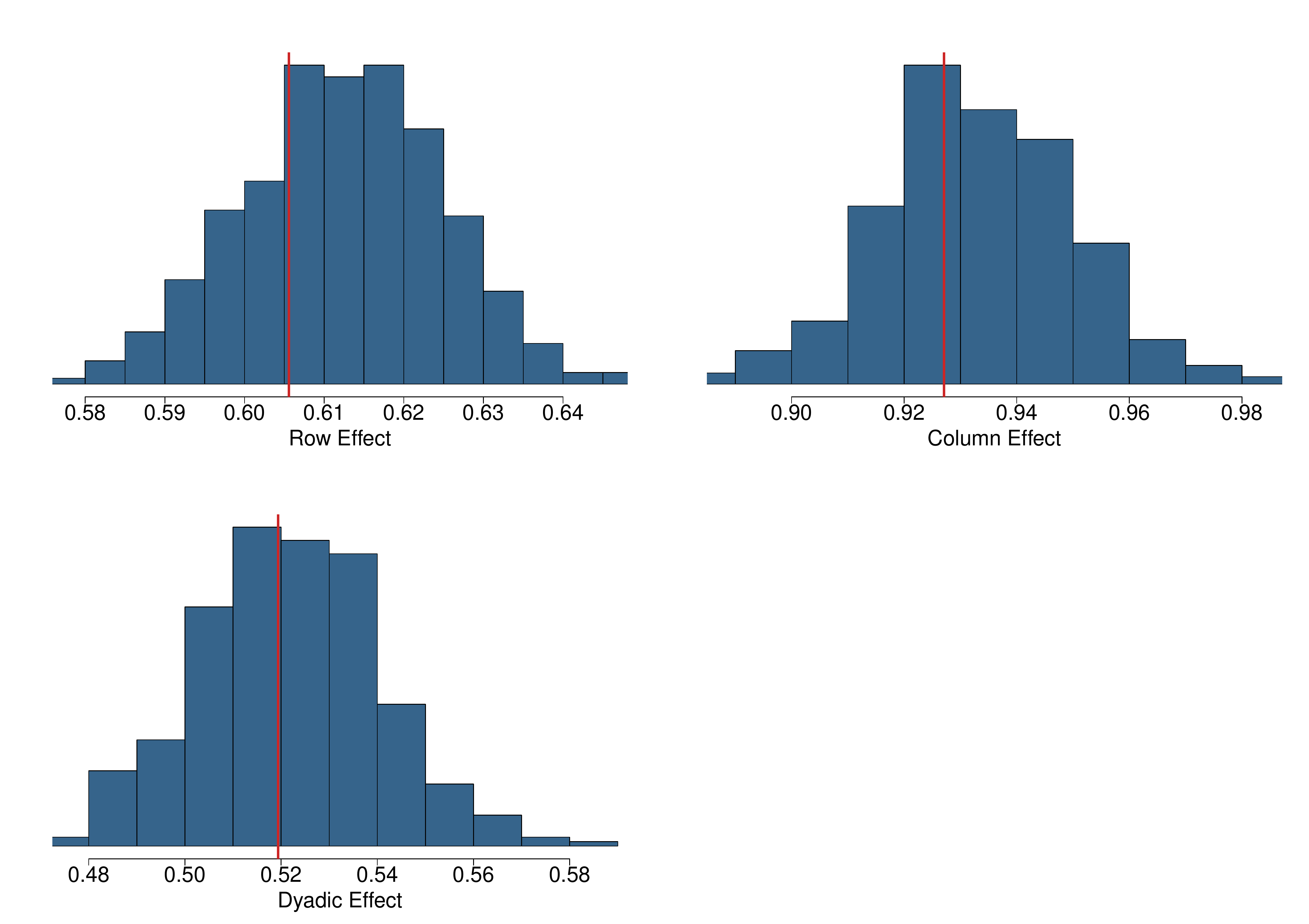}
\end{figure}

A number of checks were further conducted to measure the goodness of fit of
the model. To this aim, we compared some suitably defined summary statistics
of the observed network with the same statistics calculated from the
predicted network generated at each iteration of the MCMC algorithm. In
particular, we consider three network statistics from a given network: (1)
the standard deviation of the row means; (2) the standard deviation of the
column means; (3) the within-dyad correlation. The blue histograms in Figure %
\ref{Gof1_1} represent the posterior predictive distribution from the MCMC
inference, which are to be compared with the vertical red lines representing
the values of the statistics on the observed network. The figure shows a
good fit of the model, with the observed statistics lying at the centre of
the corresponding posterior distributions from the model.

\section{The impact of the cooperation on the hospital quality} \label{sec:ES}

As explained in the introduction we assume that
cooperation between hospitals may have a positive, negative or null effect
on overall quality in a healthcare system. In our study, quality is measured
in terms of both mortality and readmission, which are the most adopted
outcomes in the healthcare literature in order to evaluate the hospital
effectiveness. Mortality is defined by a variable assuming value 1 if the
patient dies in hospital or within 30 days after the discharge, and 0
otherwise, whereas readmission is measured by a binary variable equal to 1
if the patient is readmitted to the same hospital or to another hospital
within 45 days from a discharge and for the same clinical condition.

Using the model described in the previous section (Equation \eqref{eq7}) we
obtain the predicted transfers, $\widehat{T_{ij}}$, which, are now predicted
from a set of covariates not related to the hospital quality. In addition,
in order to avoid further problems of endogeneity, $\widehat{T_{ij}}$ is
calculated excluding the hospital random effects because these can be
related with characteristics such as the teaching status or the hospital
specialisation (cardiological hospital, neurological hospital, etc), which
can affect hospital quality. We then relate these predicted transfers with
the overall quality for the hospitals $i$ and $j$, which we measure by
\begin{equation}
W_{ij}=W_{i}+W_{j},  \label{eq_6}
\end{equation}%
when $W_{i}$ and $W_{j}$ represent the mortality/readmission of the pair of
hospitals $i$ and $j$, respectively. We decided to adopt a measure of
overall quality instead of a measure of mortality or readmissions split by
both sender and receiver hospitals because we are interested in estimating
the impact of cooperation on the overall quality of the healthcare system.
In fact, if a hospital sends a patient with a very high risk of mortality or
readmission to another hospital and the patient dies or is thereafter
readmitted, this increases the mortality and the readmission rate of the
receiver but does not impact on the quality of the pair, since the patient
would have most likely died or been readmitted in the hospital from which
he/she was transferred. Considering the overall mortality and readmissions
allows us to take into account the effect of the cooperation between
hospitals on the overall quality of both the sender and receiver hospitals,
which is at its lowest when both outcomes are being reduced.

Since the dependent variable $W_{ij}$ defined in Equation \eqref{eq_6} is
also in the form of counts, we model $W_{ij}$ via an over-dispersed Poisson
mixed effect model similar to that used in the previous section:
\begin{align}
E(W_{ij}|\cdot) = & \exp(\alpha + \xi \widehat{T_{ij}} + \beta HD_{ij} +
\theta OWN_{ij} +\delta A_{ij} + \psi F_{ij} + \phi DW_{ij} +  \notag \\
& + \vartheta Teach_{ij} + \lambda Mono_{ij} + \varrho Techno_{ij} + u_i +
u_j + \varepsilon_{ij}),  \label{eq_8}
\end{align}

where $u_k \sim N(0,\sigma_{u}^2)$, $j=1,\ldots,N$ is the hospital random
effect which is now drawn from the same distribution for the origin and
destination hospitals, since $W_{ij}$ is symmetric, and $\varepsilon_{ij}
\sim N(0,\sigma_{\varepsilon}^2)$. As expressed for the model in Equation %
\eqref{eq7}, the parameter $\sigma_{\varepsilon}^2$ captures a potential
over-dispersion in the conditional distribution of the dependent variable.

The coefficient $\xi$ in Equation \eqref{eq_8} is of interest in order to
assess the impact of cooperation on hospital quality. The model is also
scaled by the overall discharges of the hospitals' pair ($HD$), now taken as
the average of the discharges between the two corresponding hospitals since
the dependent variable is symmetric, as well as by a number of other
variables typically used in healthcare evaluations \citep{berta2016association,peluso2016pay,berta2013comparing,mascia2015effect}. In particular, we consider the patient age ($A$), the
DRG weight ($DW$) and gender ($F$), all calculated as averages of the two
connected hospitals. Moreover, we consider the hospital ownership ($OWN$), which is defined
as a dyadic variable taking values public-public, private-private,
public-private, respectively. Similarly, we also include dyadic covariates to control for the
teaching status of the hospital pair ($Teach$), their status as monospecialized or general hospitals ($Mono$), and their status as technological hospitals ($Techno$). As with ownership, these variables are defined as categorical variables, taking three possible values.

Table \ref{tab:2stage} shows the results of the model described in Equation \eqref{eq_8} using the same MCMC settings as those used for Model \eqref{eq7}. These results form the core of the paper, where we analyze the effect of
cooperation between hospitals on the quality of the healthcare system. The
analysis shows a negative and significant effect for the predicted transfers
on the health outcomes, for both mortality and readmissions. This means that
the higher is the cooperation between a pair of hospitals the higher is the
quality for the two associated hospitals, thus suggesting a positive impact
of inter-organizational cooperation to the healthcare system. In particular,
the parameters for the predicted transfers indicate that an increase of one
patient transfer produces a reduction of 1.9\% in the average mortality and
of 2.2\% in readmissions.
\begin{table}
\caption{ Modelling the effect of the network of patient flows on the
risk-adjusted mortality of the healthcare system.}
\label{tab:2stage}\centering
\begin{tabular}{lr@{\hskip .001in}lrcr@{\hskip .001in}lr}
\hline\hline
& \multicolumn{3}{c}{Mortality} &  & \multicolumn{3}{c}{Readmission} \\
\hline
& Estimate &  & Std. Err. &  & Estimate &  & Std. Err. \\
(Intercept) & 1.689 &  & 1.039 &  & 5.830 & *** & 0.913 \\
Hospital Discharges (HD) & 0.084 & *** & 0.006 &  & 0.089 & *** & 0.007 \\
DRG Weight (DW) & 0.521 & ** & 0.173 &  & 0.666 & *** & 0.183 \\
Age (A) & 0.017 & *** & 0.005 &  & -0.010 & ** & 0.005 \\
Female (F) & 0.512 &  & 0.402 &  & -0.041 &  & 0.358 \\
$\widehat{T_{ij}}$ & -0.012 & *** & 0.003 &  & -0.013 & *** & 0.003 \\
Private and Private & -0.772 & *** & 0.155 &  & -0.460 & * & 0.267 \\
Public and Private & -0.293 & *** & 0.078 &  & -0.174 &  & 0.134 \\
Technological and Technological & -0.293 & *** & 0.078 &  & -0.174 &  & 0.134
\\
Technological and Not Technological & -0.564 & *** & 0.011 &  & -0.613 & ***
& 0.009 \\
Teaching and Teaching & 0.003 &  & 0.004 &  & 0.003 &  & 0.004 \\
Teaching and Not Teaching & -0.010 &  & 0.020 &  & -0.111 & *** & 0.015 \\
Monospecialized and Monospecialized & -0.001 &  & 0.005 &  & 0.000 &  & 0.004
\\
Monospecialized and General & -0.341 & *** & 0.043 &  & 0.004 &  & 0.032 \\
\hline
$\sigma_{u}^2$ & 0.116 &  & 0.013 &  & 0.065 &  & 0.008 \\
$\sigma_{\varepsilon}^2$ & 0.074 &  & 0.004 &  & 0.055 &  & 0.002 \\
\hline\hline
\end{tabular}%
\end{table}
\begin{figure}[!h]
\caption{Posterior predictive goodness of fit for the model in equation
\eqref{eq_8}. }
\label{Gof2M_1}\centering
\includegraphics[scale = 0.35]{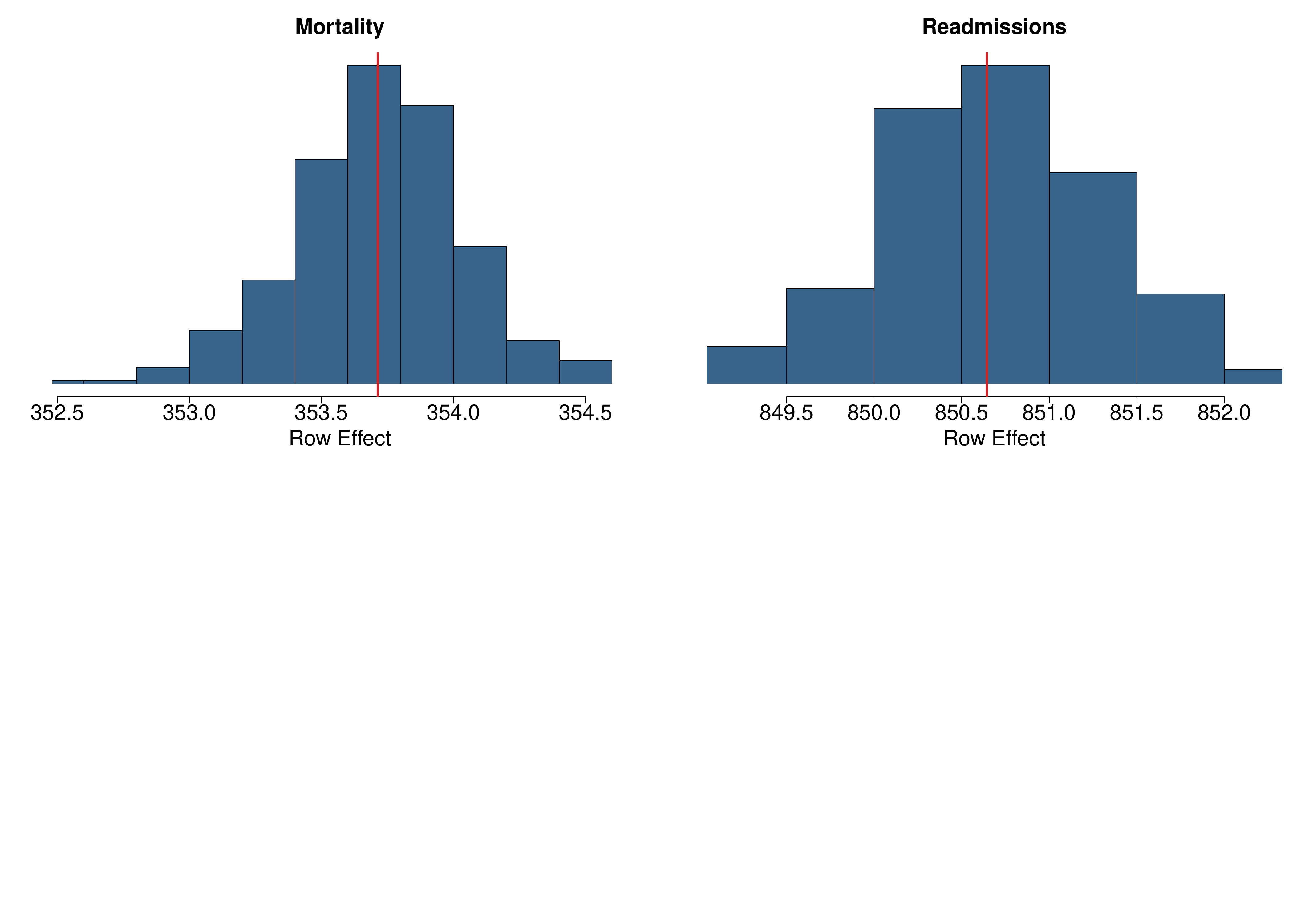}
\end{figure}
Finally, Figure \ref{Gof2M_1} shows a good fit of the models, both for the
case of mortality and readmission, with a good match between observed and
predicted summary statistics.

\begin{figure}
\caption{The effect of patient transfers on the predicted overall
mortality. The hospitals in the heatmaps are sorted by the number of
discharges and the shade of colors is defined in the log-scale, except for
the null transfers where the points are white. }
\label{fig4}\centering
\includegraphics[scale = 0.55]{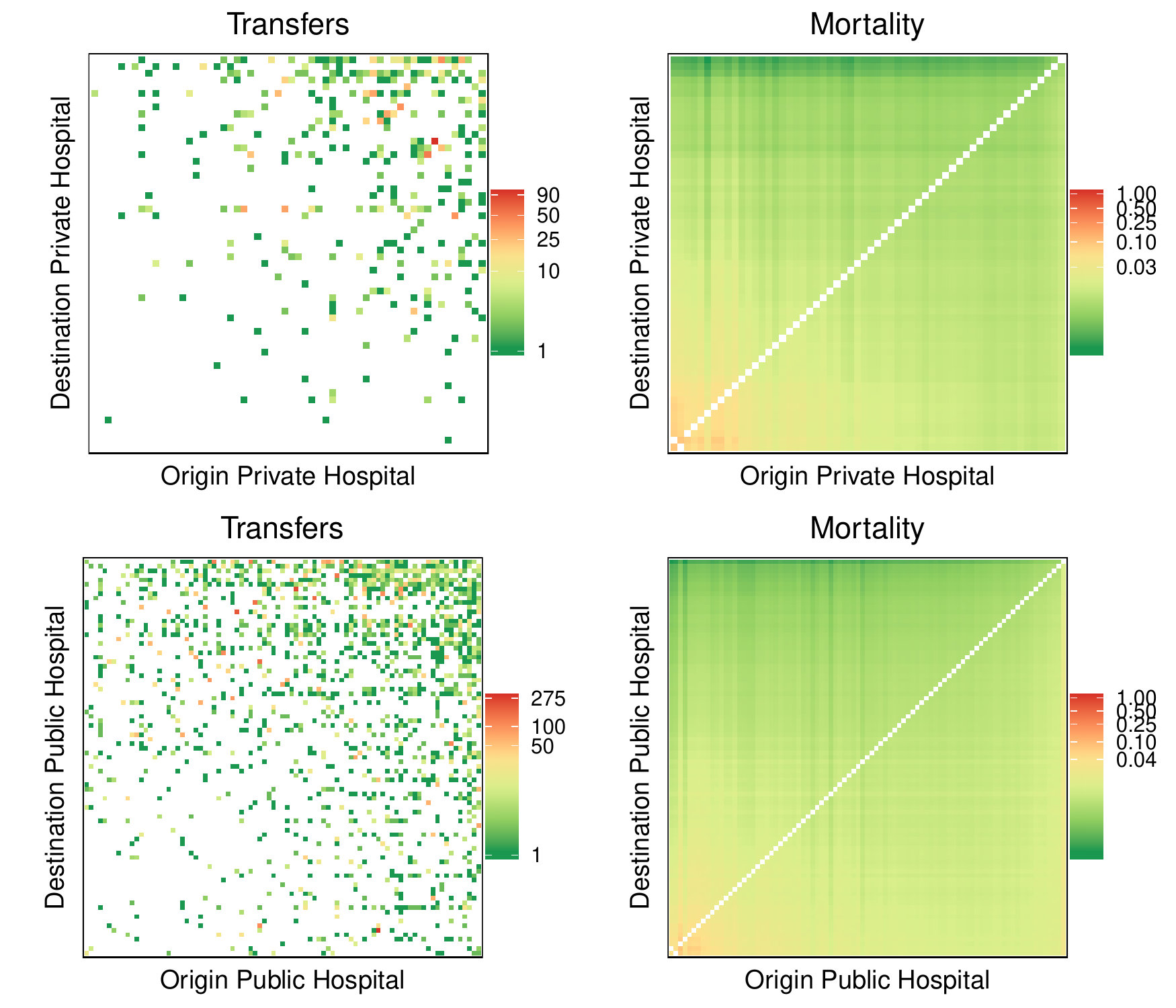}
\end{figure}

\begin{figure}
\caption{The effect of patient transfers on the predicted overall
readmission. The hospitals in the heatmaps are sorted by the number of
discharges and the shade of colors of the points is defined in the
log-scale, except for the null transfers where the points are white. }
\label{fig5}\centering
\includegraphics[scale = 0.55]{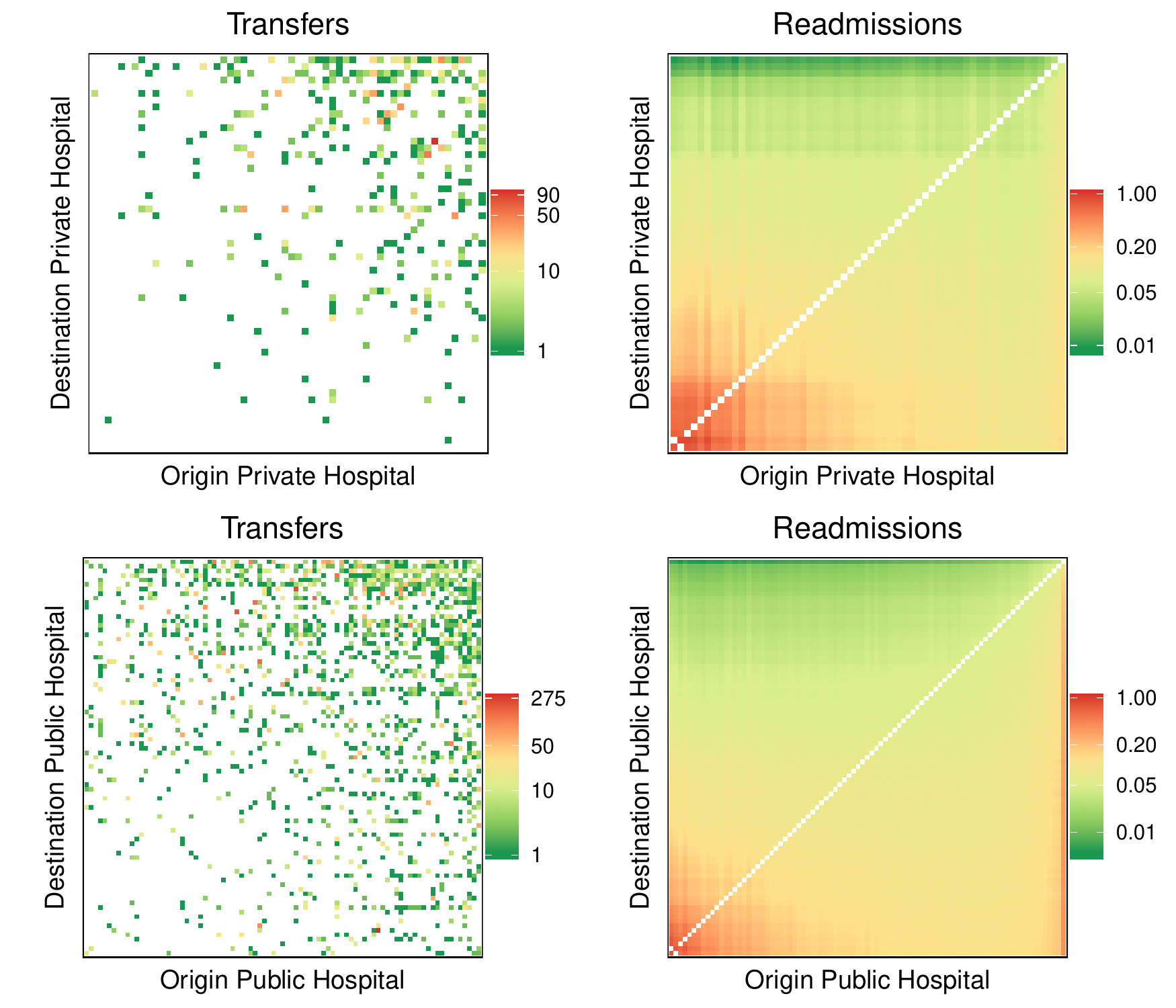}
\end{figure}

As a final step in the analysis, and considering some differences that were
previously observed between private and public hospitals, we investigate
whether the impact of cooperation on quality is different according to
hospital ownership. To this aim, we add to Equation \eqref{eq_8} an
interaction term between predicted transfers and the hospital ownerships.
The results of this analysis are presented visually in Figure \ref{fig4} and
Figure \ref{fig5}, for mortality and readmissions respectively. In both
figures, the heatmaps on the left represent the observed transfers between
hospitals' pairs sharing the same ownership (private vs private on top and
public vs public at the bottom), whereas the heatmaps on the right show the
expected mortality predicted from the model in Equation \eqref{eq_8}, scaled
by the number of discharges of the hospitals' pairs. We observe from both
analyses how public hospitals are more engaged in cooperation than private
ones, and how, in both cases, cooperation is effective in improving quality,
i.e. in reducing mortality and readmission.

\section{Discussion}

\label{sec:Discussion}

Managers from different hospitals may decide to
cooperate when treating a patient. Although we recognise that there exist
different channels of cooperation, in this paper we focus on the transfers
of patients between hospitals. We explore the impact on patient transfers
between hospitals (cooperation/network) of a set of demand-supply factors,
distance-based centrality measures that account for spatial correlation in
the data, as well as measures of quality. Our results confirm existing
findings that geographical distance plays an important role in explaining
cooperation between managers: \textit{ceteris paribus} the shorter the
distance between hospitals, the higher is the number of transfers. This
result was confirmed by the identification of communities in the network of
transfers that have a clear geographical nature. In addition, we find that
geographical centrality helps in explaining the flows of transfers: holding
all the other variables constant, the higher the number of opportunities
where to send/receive patients, the higher is the number of patients
transferred to/received by a specific hospital. Finally, we find that the intensity of cooperation between hospitals does not depend on hospital
quality. This results proves that in a healthcare system where information
on the distribution of hospital quality is not provided, managers may
exploit insight information, or have a perception of quality, in order to
engage in cooperative behaviors with other hospitals.

We then used this framework to assess, in the second stage, how such
cooperation is related to the overall clinical quality for the hospital of
origin and of destination of the patient transfer. One main challenge, when
studying this relation, is the potential endogeneity between transfers and
quality, since as we have shown in the first analysis, the decision of
transferring a patient to a specific hospital is informed, among other
things, by variables possibly correlated to the destination and origin
hospital quality. For this reason, we have also derived an exogenous measure
of transfers (cooperation) to then being able to quantify its effect on the
quality of the healthcare system. When taking care of the endogeneity of
cooperation, we find a positive relationship between hospital cooperation
and the overall quality of the connected hospitals. This is the case both
for private and public hospitals, though it is found to be more pronounced
for public hospitals.

The absence of a source of information on the quality of hospitals
accessible to all providers, such as in the form of star ratings, may
prevent some hospitals to engage with others, with some missing the
opportunity to cooperate with higher quality hospitals. This may result in a
lower degree of cooperation among hospitals and a loss of overall quality.
However, this asymmetric information may also prevent patients to choose
high quality providers \citep{berta2016association}. In other words, we
would have expected a reduction in the transfers between hospitals, had the
patients known where to be hospitalized in the first instance. Thus, the
transfers of patients can be seen as an informal mechanism in the market to
adjust ex-post for such distortion. However, even in the absence of
asymmetric information of the distribution of hospital quality, given that
some patients will not be able to choose the hospital where to be admitted
to (e.g. the urgent cases), effective cooperation between hospitals may
prove crucial in increasing the likelihood of survival for patients. In this
case, policy makers (the Lombardy Region) could design a dataset for
matching between providers in order to make cooperation more prompt and
effective.

Future work is needed in order to better assess the impact of hospital
cooperation on quality. For instance, within the category of private
hospitals there are for-profit and not for-profit organizations, with the
latter being closer to the mission of the public sector. For this reason,
future work should investigate whether this result will still hold when
considering for-profit hospitals Moreover, future work should also explore
different ways to measure the degree of cooperation between healthcare
providers. For example, it could be of interest to measure cooperation by
the scientific collaboration among professionals in different hospitals,
which can be derived by the scientific works published jointly by physicians
operating in different hospitals.

\section*{Acknowledgement}

The project was partially supported by the \emph{``European Cooperation in
Science \& Technology''} (COST) funding: action number CA15109.

\bibliographystyle{chicago}
\bibliography{biblio_Network}

\end{document}